\documentclass[preprintnumbers,twocolumn,aps,floats,nofootinbib]{revtex4-1}
\usepackage{graphicx}
\usepackage{amsmath,amssymb,amsfonts}
\usepackage{hyperref}
\usepackage{{hypcap}}
\usepackage{xspace}
\usepackage{verbatim}
\usepackage{bm}
\usepackage{color}
\usepackage{ulem}
\definecolor{light-gray}{gray}{0.85}
\newcommand{\eVdist}{\kern-0.06em}

\newcommand{\Mev}{\text{Me\eVdist V}}

\newcommand{\gev}{\:\text{Ge\eVdist V}}
\newcommand{\tev}{\:\text{Te\eVdist V}}
\newcommand{\s}{\:\text{s}}
\newcommand{\D}{\mathrm{d}}

\begin{document}
\title{Decay and Detection of a Light Scalar Boson Mixing with the Higgs}
\author{Martin Wolfgang Winkler}
\email{martin.winkler@su.se}
\preprint{NORDITA-2018-087}
\affiliation{
\vspace{4mm}
Nordita, KTH Royal Institute of Technology and Stockholm University\\
Roslagstullsbacken 23, 10 691 Stockholm, Sweden}
\begin{abstract}
The simplest extension of the standard model consists in adding one singlet scalar field which mixes with the Higgs boson. $\mathcal{O}$(GeV) masses of the new scalar carry strong motivation from relaxion, dark matter and inflation models. The decay of a GeV scalar is, however, notoriously difficult to address since, at this mass scale, the chiral expansion breaks down and perturbative QCD does not apply. Existing estimates of the GeV scalar decay rate disagree by several orders of magnitude. In this work, we perform a new dispersive analysis in order to strongly reduce these uncertainties and to resolve discrepancies in earlier results. We will update existing limits on light scalars and future experimental sensitivities which are in some cases strongly affected by the new-found decay rates. The meson form factors provided in this work, can be used to generalize our findings to non-universally coupled light scalars.
\end{abstract}
\maketitle

\section{Introduction}
Many prominent extensions of the standard model (SM) feature a gauge singlet scalar $\phi$ with a mass below or at the weak scale.
Within the relaxion mechanism~\cite{Graham:2015cka} the new scalar is introduced to cure the (little) hierarchy problem. In well-motivated dark matter models, a light scalar emerges as the mediator which links the dark and the visible sector~\cite{Pospelov:2007mp}. A light scalar appears in supersymmetric theories such as the NMSSM~\cite{Fayet:1974pd}.
It has been identified with the field driving cosmic inflation~\cite{Shaposhnikov:2006xi,Bezrukov:2009yw} and it is present in models which address the cosmological constant problem through radiative breaking of classical scale invariance~\cite{Foot:2011et}.

Through mixing with the Higgs, the light scalar inherits the Higgs couplings to SM matter reduced by a universal suppression factor. While for scalar masses around the electroweak scale, LEP and LHC constraints on extended Higgs sectors apply, rare meson decays offer a particular powerful search channel for scalars below the bottom mass threshold~\cite{Willey:1982mc}.
If the mixing is suppressed, the scalar may, however, travel a macroscopic distance before decay. In this case, searches including missing energy or displaced vertices become relevant. Present and future experimental sensitivities to a light scalar thus crucially depend on its decay rate and decay pattern. 

Since the chiral expansion breaks down shortly above the two-pion threshold, while a perturbative QCD calculation becomes reliable for masses of a few GeV, the scalar decay rate in the window $m_\phi \simeq 0.5-2\gev$ suffers from notorious uncertainties (see e.g.~\cite{Clarke:2013aya}). The problem already manifested itself when a light SM Higgs was still considered viable~\cite{Gunion:1989we}. In the late 1980s, it was realized that the form factors determining the Higgs (or general scalar) decay rate to meson final states are accessible through dispersion relations~\cite{Raby:1988qf}. Unfortunately, the two most comprehensive calculations based on this technique by Truong \& Willey~\cite{Truong:1989my} and Donoghue et al.~\cite{Donoghue:1990xh} disagree by orders of magnitude at $m_\phi\sim\text{GeV}$. It is the purpose of this work to resolve this discrepancy and to recalculate the decay rate of a light scalar to pions and kaons. Our evaluation profits from progress in the description of pion/kaon phase shift data entering the dispersive integral.

After identifying the favored parameter regions for some of the most promising SM extensions with light scalars, we will update the existing limits and future experimental sensitivities. These were previously based on varying sets of assumptions on the scalar decay. In several cases, we find the sensitivities to be substantially altered by our new-found decay rates. This holds in particular in the context of beam dump experiments which are very sensitive to the scalar decay length through the location of the detector.

\section{Standard model extensions with light scalars}\label{sec:models}
A new scalar can connect to the SM at the renormalizable level via the Higgs portal
\begin{align}
 \mathcal{L} \supset
\left(g_1 \phi + g_2 \phi^2\right) \left(H^\dagger H\right) \,.
\end{align}
Once electroweak symmetry is broken, the couplings $g_{1,2}$ induce mixing between the scalar and the Higgs. We will focus on the case where the scalar mass is considerably below the electroweak scale. In the low energy effective theory, the Higgs can then be integrated out and it arises the coupling of the new scalar to SM fermions
\begin{equation}
 \mathcal{L}\supset- \frac{s_{\theta}\,m_f}{v} \phi \bar{f}f\,,
\end{equation}
where $s_{\theta}$ denotes the sine of the Higgs-scalar mixing angle and $v$ the Higgs vacuum expectation value (vev). With regard to experimental searches, the light scalar behaves as a light version of the Higgs boson with universally suppressed couplings. In order to identify the most promising parameter space for the mixing angle, we shall briefly discuss some well-motivated SM extensions with light scalars

\subsection{Connection to Dark Matter}\label{sec:darkmatter}
New particles with a weak scale annihilation cross section have been considered among the leading dark matter candidates since --  within the thermal production mechanism -- their relic density naturally matches the observed dark matter density. The absence of a signal in direct detection experiments, however, suggests even feebler interactions between dark matter and nuclei. An appealing possibility is that dark matter resides within a dark sector of particles which do not directly feel the strong or electroweak forces~\cite{Pospelov:2007mp}. In this scenario, a scalar boson could be the mediator which communicates between dark and visible matter. In the simplest realization, dark matter is identified with a gauge singlet Majorana fermion $\chi$ which is stable due to a (discrete) symmetry and couples to the scalar via the Yukawa term~\cite{Kappl:2010qx,Schmidt-Hoberg:2013hba}
\begin{equation}
 \mathcal{L}\supset \frac{\kappa}{2}\phi\bar{\chi}\chi\,.
\end{equation}
Assuming that $m_{\chi}> m_{\phi}$, a hierarchy between the annihilation cross section and the dark matter nucleus cross section can naturally be realized: the fermions annihilate into scalars via the (unsuppressed) coupling $\kappa$, while dark matter nucleus interactions are suppressed by the mixing angle $s_{\theta}$. The annihilations cross section times relative velocity $v_{\text{rel}}$ is of the size $\sigma v_{\text{rel}}= \sigma_1 v_{\text{rel}}^2$ with~\cite{Kappl:2010qx,Winkler:2012xwa}
\begin{equation}
 \sigma_1 \simeq \frac{\kappa^4 m_\chi}{24\pi}\;\frac{9 m_\chi^4-8m_\chi^2 m_\phi^2 +2m_\phi^4}{(2m_\chi^2-m_\phi^2)^4}\;\sqrt{m_\chi^2-m_\phi^2}\,,
\end{equation}
where we assumed a vanishing trilinear scalar self-coupling for simplicity.\footnote{The general expression for the annihilation cross section for non-vanishing trilinear coupling can be found in~\cite{Winkler:2012xwa}.} Since the annihilation cross section is p-wave suppressed, strong indirect dark matter detection constraints are avoided. The fermion relic density is approximated as~\cite{Drees:2009bi}
\begin{equation}
 \Omega_\chi h^2=2.8\cdot 10^{-11}\gev^{-2}\frac{m_\chi^2}{\sqrt{g_*(T_F)}\sigma_1\,T_F^2}\,,
\end{equation}
where $g_*$ denotes the number of relativistic degrees of freedom and $T_F$ the freeze-out temperature which we take from~\cite{Steigman:2012nb}. For a given set of masses, the coupling $\kappa$ is fixed by requiring that $\Omega_\chi h^2$ matches the observed dark matter relic density. We find $\kappa=(0.03-0.05)\times\sqrt{m_\chi/\text{GeV}}$ for $m_\chi=10\:\text{MeV}-10\:\text{TeV}$.\footnote{This holds unless for very degenerate cases $m_\chi-m_\phi< 0.01\, m_\chi$.}

We have implicitly assumed a standard thermal freeze-out of the singlet fermion. This is justified if the dark sector was in thermal equilibrium with the SM bath prior to freeze-out. We, therefore, require that the thermalization rate $\Gamma_{\text{therm}}$ of the dark sector exceeds the Hubble rate of expansion $H$ at freeze-out, i.e.
\begin{equation}\label{eq:thermalization}
 \Gamma_{\text{therm}}(T_F) > H(T_F)= \sqrt{\frac{4\pi^3 g_*(T_F)}{45}}\frac{T_F^2}{\sqrt{8\pi} M_P}\,.
\end{equation}
Since $\Gamma_{\text{therm}}$ scales with $s_\theta^2$,~\eqref{eq:thermalization} puts a lower limit on the mixing angle.

At the same time, large mixing angles are excluded due to direct dark matter detection. The dark matter-nucleon cross section reads\footnote{The formula is valid for scalar masses substantially larger than the momentum transfer, i.e.\ $m_\phi\gtrsim 100\:\text{MeV}$.}~\cite{Schmidt-Hoberg:2013hba}
\begin{equation}
 \sigma_n\simeq \frac{4\mu_\chi^2}{\pi}\left(\frac{s_{\theta}\kappa}{2 v m_\phi^2}\right)^2 \! m_n^2 \left(f^n_u+f^n_d+f^n_s+\frac{6}{27} f_G\right)^2
\end{equation}
with $m_n$ denoting the nucleon mass and $\mu_\chi$ the reduced mass of the dark matter-nucleon system. The scalar coefficients $f_{u,d,s}^n$ and $f_G$ define the quark and gluon content of the nucleon for which we employ the standard values given in~\cite{Belanger:2013oya}. The dark matter direct detection constraints can now be mapped into the scalar mass-mixing plane. Besides the constraints of XENON1T~\cite{Aprile:2018dbl}, we also include those of CRESST-III~\cite{Petricca:2017zdp} and DarkSide-50~\cite{Agnes:2018ves} which dominate at $m_\chi\lesssim 5\gev$. 

Since the most conservative (weakest) bounds are obtained if $\chi$ is just slightly heavier than $\phi$, we fix $m_\chi=1.1\,m_\phi$. In this case, the thermalization rate is dominated by the inverse decay of the scalar~\cite{Evans:2017kti} and we have to apply~\eqref{eq:thermalization} with $\Gamma_{\text{therm}}\simeq\Gamma_\phi$. As shown in figure~\ref{fig:limits}, the parameter space, where thermalization and direct detection constraints can simultaneously be satisfied spans several orders of magnitude in $s_\theta$. Further experimental constraints on this window will be discussed in section~\ref{sec:constraints}. 

\subsection{Relaxion}\label{sec:relaxion}

The relaxion mechanism constitutes a dynamical solution to the (little) hierarchy problem of the standard model~\cite{Graham:2015cka}. It  provides another motivation for the existence of a light scalar boson. While the phenomenology of Higgs-relaxion mixing has been comprehensively studied~\cite{Choi:2016luu,Flacke:2016szy}, we wish to include the additional possibility of a low inflationary Hubble scale $H_I$. 

The evolution of the relaxion $\phi$ reduces the initially large Higgs boson mass $M\gg v$ to the observed mass $m_h=\mathcal{O}(v)$. This is achieved via the potential\footnote{We neglect an $\mathcal{O}(1)$ coefficient in front of the $g\,M^3$ term which does not play a role for the following discussion.}
\begin{equation}\label{eq:relaxionpotential}
V = (M^2 - gM \phi) \,h^2 - g\,M^3 \phi - \Lambda^2 \,h^2\, \cos\left(\tfrac{\phi}{f}\right)+\lambda\,h^4\,,
\end{equation}
where $g$ is a dimensionless coupling and $h$ denotes the neutral component of the Higgs doublet. Since the relaxion settles in a CP breaking minimum, it is not identified with the QCD axion in the basic model. Instead, the periodic potential may stem from the instantons of a new strongly coupled gauge group~\cite{Graham:2015cka}.\footnote{For concreteness, we assumed that the new strongly coupled sector does not break electroweak symmetry such that odd powers of $h$ are absent in front of the cosine. The phenomenology is, however, hardly sensitive to this assumption (see~\cite{Flacke:2016szy}).} The scale $\Lambda$ must not exceed the electroweak scale since, otherwise, the Higgs vev is driven up to $\Lambda$. This constraint also ensures that a constant term in front of the cosine, which is generated by closing the Higgs loop, is sufficiently suppressed and does not trap the relaxion before electroweak symmetry breaking~\cite{Choi:2016luu,Flacke:2016szy}.

The relaxion slowly rolls down its potential and, at $\phi \sim M/g$ triggers electroweak symmetry breaking. As soon as the Higgs field is displaced, the cosine term induces wiggles on the relaxion potential which ultimately stop its motion. The required dissipation mechanism is provided by the Hubble friction of inflation. If $H_I$ exceeds a critical value $H_{I,c}\sim\sqrt{g\,M^3/f}$, the relaxion immediately stops in one of its first minima. Otherwise, it continues rolling and later settles in one of the steeper minima, further down the potential. The difference between both cases manifests in the phase factor 
\begin{equation}
 \sin\left(\frac{v_\phi}{f}\right)\sim \text{Min}\left(1,\,\frac{H_I^2\,f}{g\,M^3}\right)\,,
\end{equation}
where we introduced the relaxion vev $v_\phi$. The sine is of order unity if $H_I>H_{I,c}$, while it can be substantially suppressed for a low inflationary scale. The Higgs vev emerges as
\begin{equation}
 v^2  \simeq \frac{g M^3 f}{\Lambda^2\,\sin\left(\frac{v_\phi}{f}\right)}\,.
\end{equation}
Validity of the effective theory~\eqref{eq:relaxionpotential} without further light degrees of freedom requires $f \gg v \gtrsim \Lambda$. This implies that the relaxion is lighter than the Higgs and the mixing effect on $m_h$ is negligible. The relaxion mass\footnote{More precisely, we are referring to the mass of the relaxion-like scalar mass eigenstate.} and the Higgs-relaxion mixing angle can be approximated as~\cite{Flacke:2016szy}
\begin{align}
 m_\phi^2 &\simeq \frac{\Lambda^2 v^2}{2 f^2} \left[\cos\left(\frac{v_\phi}{f}\right)-\frac{2\Lambda^2}{m_h^2}\sin^2\left({\frac{v_\phi}{f}}\right)\right]\,,\nonumber\\
 s_\theta& \simeq \frac{\Lambda^2 v}{f m_h^2}\sin\left(\frac{v_\phi}{f}\right)\,.
\end{align}
The relaxion couples to SM matter via its Higgs admixture and via pseudoscalar couplings which are generically present but model-dependent. Requiring that the mixing-induced couplings dominate leads to the constraint $\sin(v_\phi/f)\gtrsim 1/(16\pi^2)$.\footnote{For $\sin(v_\phi/f)\gtrsim 1/(16\pi^2)$, the CP violating scalar relaxion couplings can still dominate since pseudoscalar couplings may suffer additional loop suppression~\cite{Flacke:2016szy}. We note that viable relaxion models with smaller $\sin(v_\phi/f)$ may exist. The constraint, however, singles out the parameter region in which the relaxion can be described as a minimal singlet scalar mixing with the Higgs.} The resulting theory exclusion on the parameter space (requiring also $f>v$) is depicted in figure~\ref{fig:limits}. Compared to~\cite{Choi:2016luu,Flacke:2016szy}, we obtain a larger relaxion window since suppression of $s_\theta$ by small $H_I$ has not been considered in these references.

\section{Scalar Decay Rates}\label{sec:scalardecay}

It is straightforward to evaluate the scalar decay rates into leptonic final states. One finds
\begin{equation}\label{eq:mudecay}
 \Gamma (\phi\rightarrow \bar{\ell}\ell) \equiv \Gamma_{\bar{\ell}\ell} = \frac{s_\theta^2\,G_F\,m_\phi}{4\sqrt{2}\,\pi}\, m_\ell^2\,\beta_\ell^3\;,
\end{equation}
with $\ell=e,\mu,\tau$. Here, $G_F$ denotes the Fermi constant and $\beta_\ell = \sqrt{1-4m_\ell^2/m_\phi^2}$ 
the velocity of the final state leptons. Hadronic decay rates require a more careful treatment due to the strong final state interactions. This holds in particular, if the scalar mass resides in the vicinity of the $f_0(980)$ resonance.

\subsection{Status of Hadronic Decay Rates}
\begin{figure}[htp]
\begin{center}
 \includegraphics[width=8.5cm]{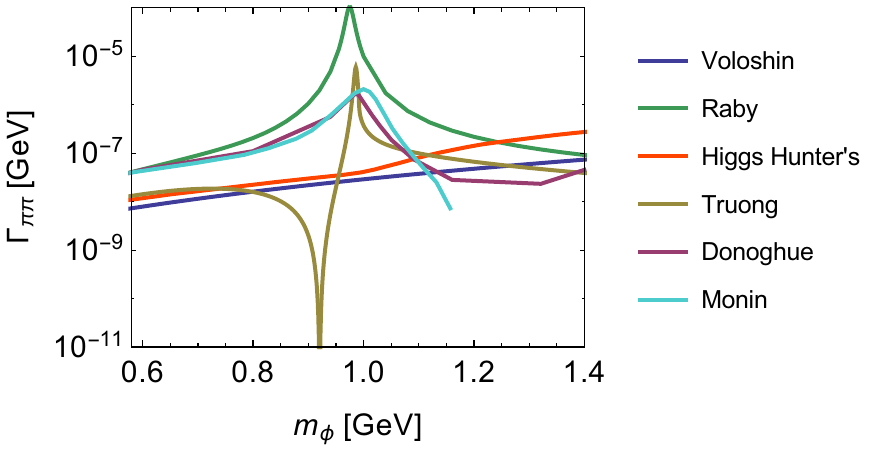}\\
\end{center}
\caption{Evaluations of the light scalar decay rate to pions by Voloshin~\cite{Voloshin:1985tc}, Raby \& West~\cite{Raby:1988qf}, the Higgs Hunter's Guide~\cite{Gunion:1989we}, Truong \& Willey~\cite{Truong:1989my}, Donoghue et al.~\cite{Donoghue:1990xh} and Monin et al.~\cite{Monin:2018lee}. In this figure $s_\theta$ has been set to unity.} 
\label{fig:oldrates}
\end{figure}

Figure~\ref{fig:oldrates} shows that different evaluations of the scalar decay rate to pions disagree by several order of magnitude at $m_\phi\sim\text{GeV}$. The result of Voloshin was obtained at leading order in chiral perturbation theory (ChPT)~\cite{Voloshin:1985tc}. In the `Higgs Hunter's guide' the perturbative spectator model is extrapolated into the non-perturbative regime. Quark masses were adjusted such as to (approximately) reproduce Voloshin's decay rate at low mass~\cite{Gunion:1989we}. Both evaluations are frequently used to describe GeV scalar decays although they do not apply to this mass range due to its proximity to the chiral symmetry breaking scale. Raby \& West~\cite{Raby:1988qf} introduced the use of dispersion relations to access the GeV regime and predicted a huge enhancement of the scalar decay rate to pions close to the $f_0(980)$ resonance. However, they treated $f_0(980)$ as an elastic $\pi\pi$-resonance which leads to an overestimation of the rate. A full two-channel analysis including $KK$ and $\pi\pi$ was finally performed by Truong \& Willey~\cite{Truong:1989my} and Donoghue et al.~\cite{Donoghue:1990xh}. Unfortunately, their results are incompatible with one another. Monin et al.~\cite{Monin:2018lee} recently performed a modified one-channel analysis in order to provide an analytic expression for $\Gamma_{\pi\pi}$ in terms of the $\pi\pi$-scattering phase. Since free parameters were chosen with the purpose of reproducing the rate of Donoghue et al., it was not meant as a test of previous results. A calculation of the hadronic decay rates in an independent two-channel dispersive analysis is still missing. It will be performed in the next sections, before matching the result to the perturbative spectator model at higher mass.

\subsection{Chiral Perturbation Theory}

We first consider scalar masses below the charm threshold. The Lagrangian describing the interaction of the scalar with light quarks ($u,d,s$) and gluons reads
\begin{align}
  \mathcal{L} &\supset  s_{\theta}\frac{\phi}{v} \left( \frac{3\alpha_s}{12\pi}G^{a}_{\mu\nu}G^{a\mu\nu}  - m_u\bar{u}u - m_d\bar{d}d - m_s\bar{s}s\right)\nonumber\\
&= - s_{\theta}\frac{\phi}{v} \left( \frac{2}{9}\Theta_\mu^\mu +  \frac{7}{9} \left(m_u\bar{u}u+m_d\bar{d}d+m_s\bar{s}s\right)\right)\,,
\end{align}
where the effective coupling to gluons origins from heavy quark ($c,b,t$) loops. In the second step, we used the trace identity 
\begin{equation}
 \Theta_\mu^\mu = -\frac{9\alpha_s}{8\pi} G^{a}_{\mu\nu}G^{a\mu\nu}  + m_u\bar{u}u + m_d\bar{d}d + m_s\bar{s}s\,,
\end{equation}
of the energy-momentum tensor which results from the conformal anomaly~\cite{Crewther:1972kn,Chanowitz:1972vd}. The decay rates of the scalar into pion and kaon pairs read
\begin{align}
\Gamma_{\pi\pi}&= \frac{3s_{\theta}^2\, G_F}{16\sqrt{2}\pi m_\phi}\beta_\pi \left|\tfrac{7}{9}\Gamma_\pi + \tfrac{7}{9}\Delta_\pi+ \tfrac{2}{9}\Theta_\pi\right|^2\,,\nonumber\\
\Gamma_{KK}&= \frac{s_{\theta}^2\,G_F}{4\sqrt{2}\pi m_\phi}\beta_K \left|\tfrac{7}{9}\Gamma_K + \tfrac{7}{9}\Delta_K+ \tfrac{2}{9}\Theta_K\right|^2\,,
\end{align}
where we introduced the form factors
\begin{align}
 \Gamma_\pi &= \langle \pi\pi | m_u \bar{u} u+m_d \bar{d} d|0\rangle\,,\nonumber\\
 \Delta_\pi &= \langle \pi\pi | m_s \bar{s} s|0\rangle\,,\nonumber\\
 \Theta_\pi &= \langle \pi\pi | \Theta_\mu^\mu|0\rangle\,,
\end{align}
for pions and analogous for kaons. The pion form factors have been determined to lowest order in ChPT in~\cite{Voloshin:1985tc}. A ChPT calculation of the kaon form factors may seem pointless since the scalar decay to kaons only opens in the regime, where chiral symmetry is strongly broken. However, the low-momentum kaon form factors will later define the matching conditions for the dispersive analysis. Therefore, we briefly outline the computation using the (strangeness-conserving part of the) 3-flavor chiral Lagrangian which reads\footnote{An analogous determination of the kaon form factors can be found in~\cite{Donoghue:1990xh}. For a review on the application of ChPT techniques to Higgs physics, see~\cite{Dawson:1989yh,Gunion:1989we}.}
\begin{equation}
 \mathcal{L} = \frac{1}{4}f_\pi \text{Tr}\partial_\mu\Sigma\partial^\mu\Sigma^\dagger + \frac{1}{2} f_\pi^2 \left( \text{Tr} \mu M \Sigma^\dagger +\text{h.c.}\right)\,,
\end{equation}
with
\begin{equation}
 \Sigma = \exp\left\{ \frac{\sqrt{2} i}{f_\pi} 
 \left(\begin{smallmatrix}
  \frac{\pi^0}{\sqrt{2}} + \frac{\eta}{\sqrt{6}} & \pi^+ & K^+\\
  \pi^- & -\frac{\pi^0}{\sqrt{2}}+\frac{\eta}{\sqrt{6}} & K^0\\
  K^- & \bar{K}_0 & -\frac{2\eta}{\sqrt{6}}
   \end{smallmatrix}\right)
\right\}
\end{equation}
and $M=\text{diag}(m_u,m_d,m_s)$. Here $f_\pi$ denotes the pion decay constant. The mass parameters in the chiral Lagrangian are related to the physical meson masses as
\begin{align}
 m_\pi^2&= \mu (m_u+m_d)\,,\nonumber\\
 m_{K^0}^2&= \mu(m_d + m_s)\,,\nonumber\\
 m_{K^\pm}^2&= \mu(m_u + m_s)\,.
\end{align} 
One can now use the Feynman-Hellmann theorem~\cite{Hellmann:1937,Feynman:1939zza} $m_q\bar{q}q = -m_q \partial \mathcal{L}/ \partial m_q$ and the trace of the energy-momentum tensor
\begin{equation}
 \Theta_\mu^\mu = \tfrac{f_\pi}{2} \text{Tr}\,\partial_\mu\Sigma\partial^\mu\Sigma^\dagger - g_\mu^\mu \mathcal{L}\,,
\end{equation}
to evaluate the form factors at lowest order in the chiral expansion (denoted by the superscript $0$). One finds
\begin{align}\label{eq:formfactors}
 \Gamma_\pi^0 & = m_\pi^2\,, & \Gamma_K^0 &= \tfrac{1}{2} m_\pi^2\,,\nonumber\\
 \Delta_\pi^0 &=  0\,, & \Delta_K^0 &= m_K^2-\tfrac{1}{2}m_\pi^2\,,\nonumber\\
 \Theta_\pi^0 &=  s + 2m_\pi^2\,, & \Theta_K^0 &=  s + 2m_K^2\,,
\end{align} 
where we set $m_u = m_d$. The form factors have to be evaluated at $\sqrt{s}=m_\phi$. Higher orders are suppressed by powers of the chiral symmetry breaking scale $\Lambda_\chi\sim 1\:\text{GeV}$. The lowest order does, hence, not provide a realistic estimate of the form factors for $m_\phi \gtrsim 0.5\gev$.

\subsection{Dispersive Analysis}

Fortunately, form factors at higher mass are accessible through dispersion relations. These employ analyticity and unitarity conditions without relying on any details of the microscopic interaction theory. For $\sqrt{s}\lesssim 1.3\gev$ a two-channel approximation in terms of $\pi\pi$ and $KK$ can be applied. This is because scalar decays are controlled by the $f_0(980)$ resonance at $\sqrt{s}\sim\text{GeV}$ which mainly couples to these states~\cite{Patrignani:2016xqp}. At even lower mass, $\pi\pi$ is the only relevant decay channel due to kinematics.

We define $F=(F_\pi,\frac{2}{\sqrt{3}}F_K)$ ($F=\Gamma,\,\Delta,\,\Theta$), where the Clebsch-Gordan coefficient occurring in the isoscalar projection of the $\pi\pi$ state has been absorbed into the definition of $F$~\cite{Donoghue:1990xh}. Below the kaon threshold, the phase of the pion form factors coincides with the isoscalar s-wave $\pi\pi$ phase shift according to Watson's theorem~\cite{Watson:1954uc}. Its generalization to two channels is expressed in form of the unitary relation 
\begin{equation}\label{eq:unitary}
 \text{Im}F_i = T_{ij}^* \:\beta_j\, F_j \,\theta(s-4m_j^2) 
\end{equation}
with $\beta_{1,2}=\beta_{\pi,K}$. The (isoscalar s-wave projection of the) $T$-matrix for $\pi\pi,KK\rightarrow \pi\pi,KK$ scattering is parameterized in terms of two phases $\delta$, $\psi$ and an inelasticity parameter $g$
\begin{equation}
 T=\begin{pmatrix}
    \frac{\eta\:e^{2i\delta}-1}{2i\beta_\pi} &
    g \,e^{i \psi}\\
    g \,e^{i \psi}&
\frac{\eta\:e^{2i(\psi-\delta)}-1}{2i\beta_K}  
   \end{pmatrix}
   \;,
\end{equation}
where
\begin{equation}
\eta=\sqrt{1-4\,\beta_\pi\beta_K \,g^2 \:\theta(s-4 m_K^2)}\,.
\end{equation}
The parameters of the $T$ matrix are efficiently determined by invoking $\pi\pi,\,KK$ scattering data and theoretical constraints in form of the Roy-Steiner equations. We extract the phases and inelasticity parameter from the analysis of Hoferichter et al.~\cite{Hoferichter:2012wf} which incorporates earlier results~\cite{Caprini:2011ky,Buettiker:2003pp}. Above $\sqrt{s_0}=1.3\gev$, the correct asymptotic behavior of the $T$-matrix is ensured by guiding $\delta,\,\psi$ smoothly to $2\pi$ according to eq.\ (41) in~\cite{Moussallam:1999aq}. We have verified that form factors at $\sqrt{s}<\sqrt{s_0}$ are rather insensitive to the particular function by which the phases approach their asymptotic values. Above $\sqrt{s_0}$ the form factors obtained from the two-channel analysis are anyway less trustable since further channels like $4\pi,\,\eta\eta$ become relevant.

Form factors satisfying the unitary relation~\eqref{eq:unitary} can be expressed as~\cite{Muskhelishvili1953,Omnes:1958hv}
\begin{equation}\label{eq:omnessolution}
 F= \begin{pmatrix}
 \Omega_{11} &\Omega_{12} \\    
 \Omega_{21} &\Omega_{22}
    \end{pmatrix}
    \begin{pmatrix}
 P_1 \\    
 P_2
    \end{pmatrix}\,,
\end{equation}
where $P_{1,2}$ are polynomials and $(\Omega_{11},\Omega_{21})$, $(\Omega_{12},\Omega_{22})$ are the two linear independent solution-vectors fulfilling the dispersion relation
\begin{equation}
 \text{Re}F(s) = \frac{1}{\pi}\mathop{\,\text{--}\hspace{-3.4mm}\int}\limits_{4m_\pi^2}^{\infty} \D s^\prime \frac{\text{Im}F(s^\prime)}{s^\prime-s}\,.
\end{equation}
The $\Omega_{ij}$ (which are found as described in~\cite{Moussallam:1999aq}) are conveniently normalized such that $\Omega_{11}(0)=\Omega_{22}(0)=1$, $\Omega_{12}(0)=\Omega_{21}(0)=0$. 

\begin{figure}[t]
\flushright   
 \includegraphics[width=8cm]{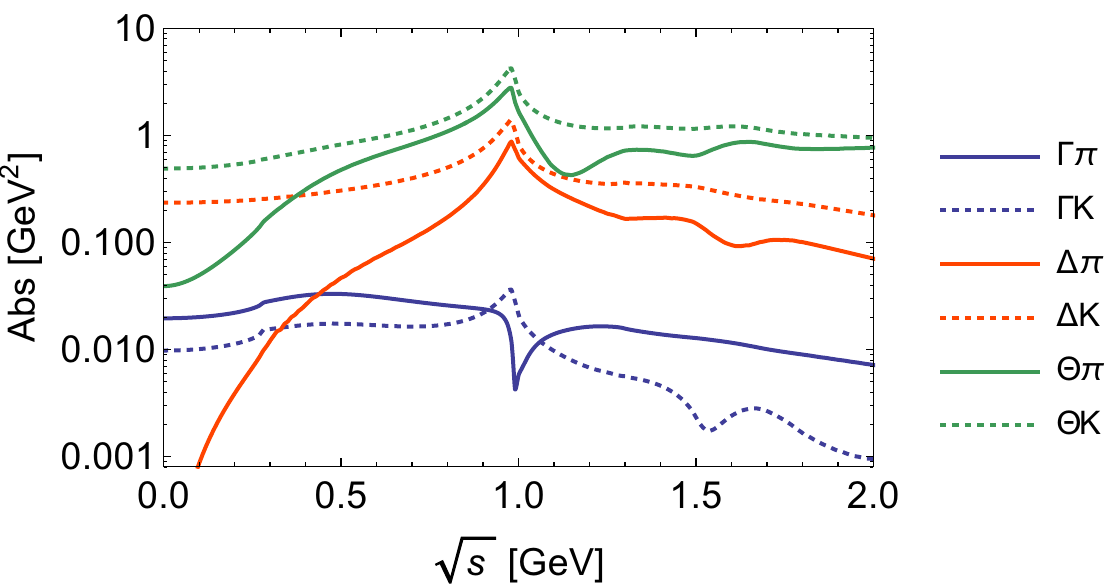}$\;\;\;\;$\\
 \includegraphics[width=7.5cm]{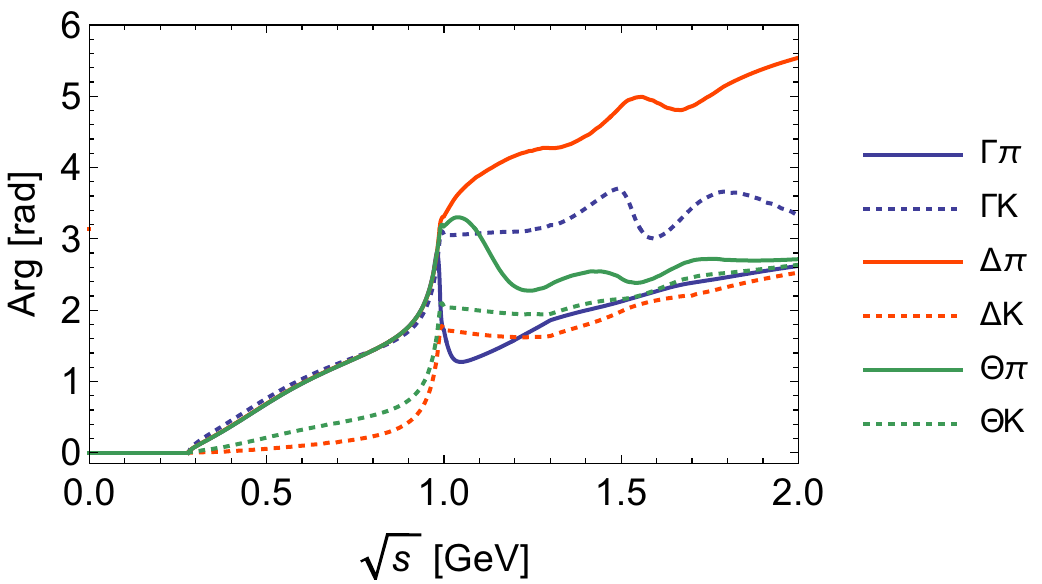}$\;\;\;\;$
\caption{Modulus (upper panel) and phase (lower panel) of the pion and kaon form factors.}
\label{fig:formfactors}
\end{figure}

\begin{figure}[htp]
\begin{center}
 \includegraphics[width=7.5cm]{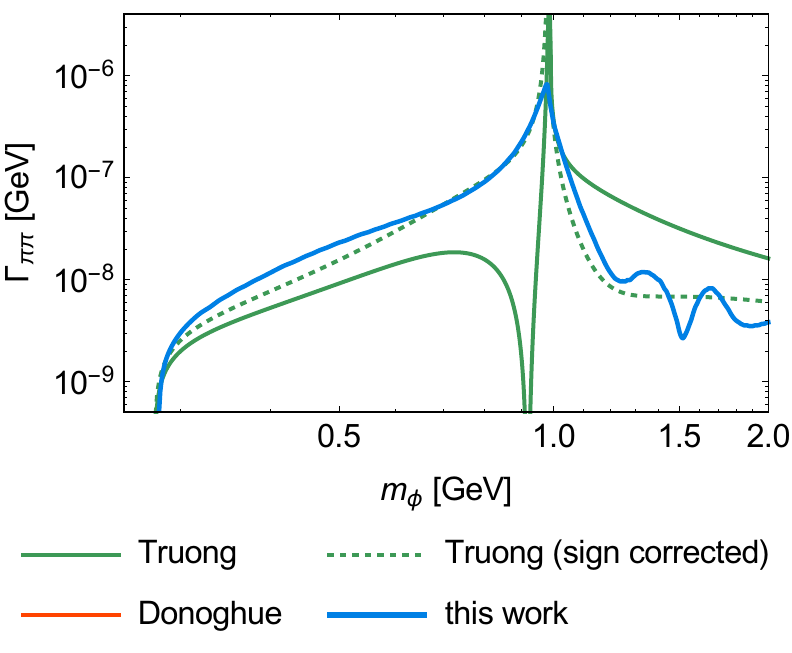}\\
 \includegraphics[width=7.5cm]{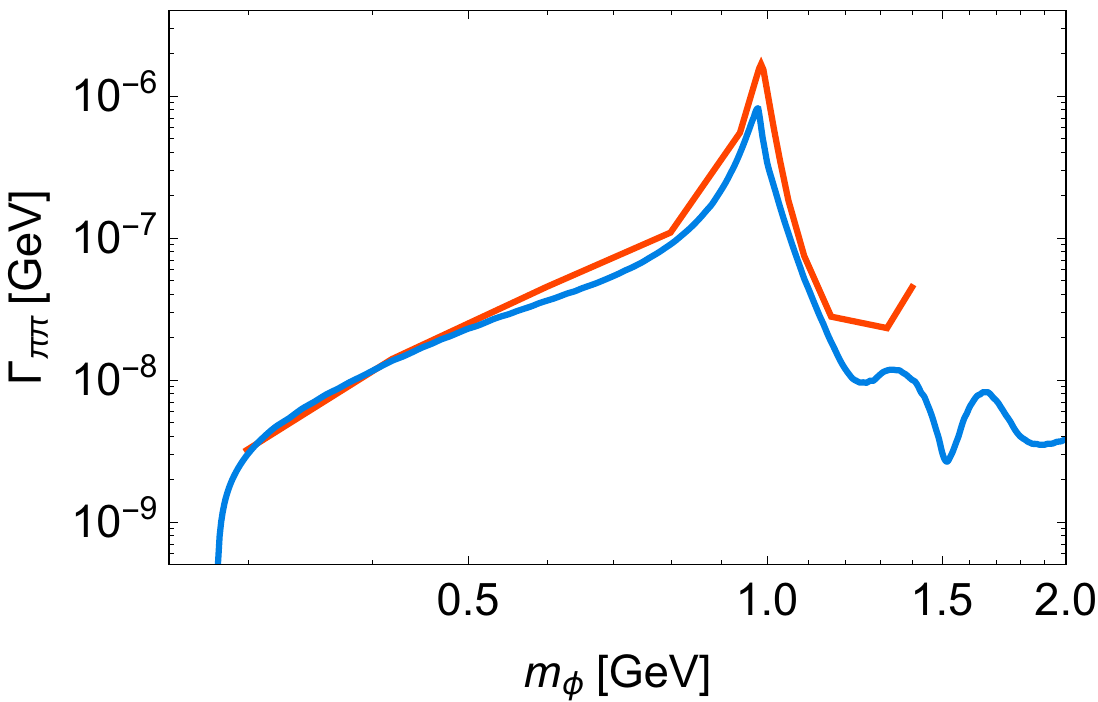}
\end{center}
\caption{Light scalar decay rate into pions from this work, from Truong \& Willey~\cite{Truong:1989my} and from Donoghue et al.~\cite{Donoghue:1990xh}. We also show Truong \& Willey's decay rate after correcting a sign error in their $T$-matrix parameterization (see text).} 
\label{fig:donwil}
\end{figure}

The form factors $\Gamma_i,\,\Delta_i$ are expected to vanish at high energy due to the composite nature of mesons. Since $\Omega_{ij}\propto s^{-1}$ for large $s$, the polynomial prefactors in~\eqref{eq:omnessolution} need to be constants. Their values can be determined by matching~\eqref{eq:omnessolution} to the lowest order result in chiral perturbation theory~\eqref{eq:formfactors} at $s=0$. In the case of the energy-momentum form factors, Lorentz-invariance and four-momentum conservation require the structure~\cite{Donoghue:1991qv}
\begin{equation}
 \Theta_i = \frac{3}{2} s\: \Theta_{S,i} + \left(2m_i^2-\frac{s}{2}\right) \Theta_{T,i}\qquad (i=\pi,\,K)\,,
\end{equation}
where $\Theta_{S,i}$ and $\Theta_{T,i}$ refer to the scalar and tensor parts of $\Theta_i$. In order to match the chiral result at $s=0$, one needs to require that $\Theta_{S,i},\,\Theta_{T,i}$ (rather than $\Theta_i$) vanish asymptotically (see also~\cite{Donoghue:1990xh}). We thus obtain
\begin{align}
 \Gamma_\pi&= m_\pi^2\left(\Omega_{11} + \frac{1}{\sqrt{3}} \Omega_{12}\right)\,,\nonumber\\
 \Delta_\pi&= \frac{2}{\sqrt{3}}\left(m_K^2-\frac{m_\pi^2}{2}\right)\Omega_{12}\,,\nonumber\\
 \Theta_\pi&= \left(2 m_\pi^2 + p s\right)\Omega_{11} +  \frac{2}{\sqrt{3}}\left(2m_K^2 + q s\right)\Omega_{12}\,,\nonumber\\ 
  \Gamma_K&= \frac{m_\pi^2}{2}\left(\sqrt{3}\,\Omega_{21} + \Omega_{22}\right)\nonumber\,,\nonumber\\
 \Delta_K&= \left(m_K^2-\frac{m_\pi^2}{2}\right)\Omega_{22}\,,\nonumber\\
  \Theta_K&= \frac{\sqrt{3}}{2}\left(2 m_\pi^2 + p s\right)\Omega_{21} +  \left(2m_K^2 + q s\right)\Omega_{22}\,,
\end{align}
where we introduced
\begin{align}
 p&=1-2 m_\pi^2\,\Omega_{11}^\prime(0)-\frac{4 m_K^2}{\sqrt{3}}\,\Omega_{12}^\prime(0)\,,\nonumber\\
 q&=1-\sqrt{3} m_\pi^2\, \Omega_{21}^\prime(0)-2 m_K^2\,\Omega_{22}^\prime(0)\,.
\end{align}
Numerically, we find $p=0.73$ and $q=0.52$. In figure~\ref{fig:formfactors} we depict the resulting pion and kaon form factors. The corresponding scalar decay rates to pions and kaons agree reasonably well with the result of Donoghue et al.~\cite{Donoghue:1990xh} (see figure~\ref{fig:donwil}). Differences reside within a factor of $\sim 3$ and follow from our updated phase shift input~\cite{Hoferichter:2012wf}. The decay rates found by us are, however, incompatible with those in~\cite{Truong:1989my}. The reason for the discrepancy is indeed a sign error in Truong \& Willey's parameterization of the $T$-matrix. Their choice leads to a negative sign of $T_{12}$ at low energy which is inconsistent with ChPT~\cite{Donoghue:1990xh}. In figure~\ref{fig:donwil} we also depict the decay rate after flipping the sign of their parameter $\lambda$. It can be seen that this correction puts Truong \& Willey's rate into qualitative agreement with our result. 

\subsection{Perturbative Spectator Model}
\begin{figure*}[htp]
\begin{center}
 \includegraphics[width=12cm]{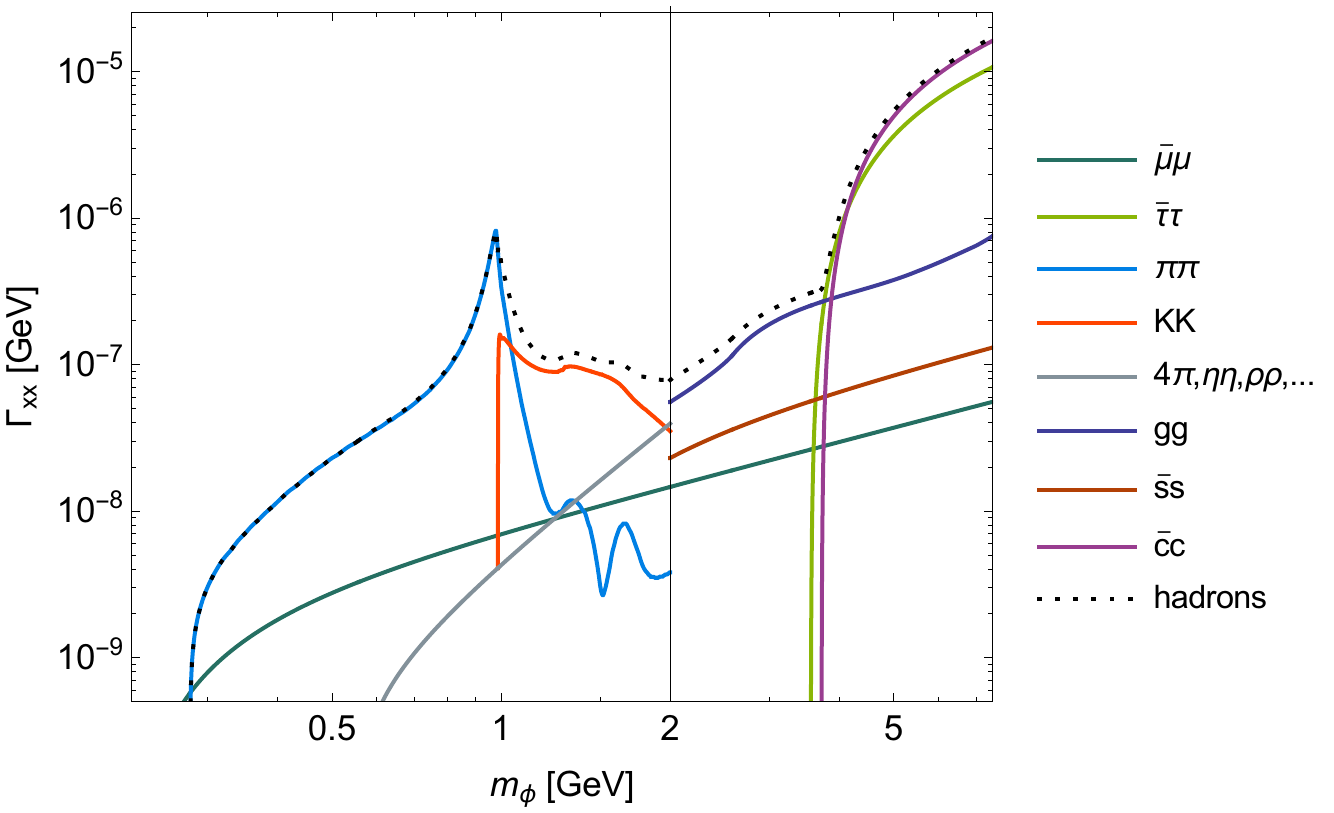}
 \end{center}
\caption{Hadronic and leptonic decay rates of a light scalar mixing with the Higgs. The decay rates scale with $s_\theta^2$ which was set to unity in this plot.} 
 \label{fig:decayrates}
\end{figure*}

We now turn to the hadronic decays at higher energy, where the perturbative spectator model can be applied. The decay rates to quarks are given as\footnote{We set $m_s=95\Mev$, $m_c=1.3\gev$~\cite{Patrignani:2016xqp} and neglect the tiny decay rate into $u$, $d$ quarks.}
\begin{align}\label{eq:branching}
  \Gamma_{\bar{\ell}\ell}\: : \:\Gamma_{\bar{s}s}\: : \:\Gamma_{\bar{c}c}
= \: m_\ell^2\,\beta_\mu^3 \: : \: 3\,m_s^2\beta_K^3\: : \:
3 m_c^2\beta_D^3
\end{align}
and analogous for the $\bar{b}b$-channel. The kinematic threshold is set by the lightest meson containing an $s$ or $c$ quark respectively~\cite{Gunion:1989we}. In addition, we need to consider the loop-induced decay rate into gluon pairs~\cite{Spira:1995rr}
\begin{equation}
  \Gamma_{gg} = \frac{s_{\theta}^2\,\alpha_s^2\, m_\phi^3}{32\pi^3 v^2} \left| \sum_{\text{quarks}} \frac{x_i +(x_i-1) f(x_i)}{x_i^2}\right|^2\,,
\end{equation}
with $x_i= m_\phi^2 / (4 m_i^2)$ and
\begin{equation}
f(x) = \begin{cases} 
\text{arcsin}^2\sqrt{x}\,,\quad x\leq 1\\
-\frac{1}{4} \left( \log\frac{1+\sqrt{1-1/x}}{1-\sqrt{1-1/x}} - \text{i} \pi  \right)^2\,,\quad x>1\,.
\end{cases}
\end{equation}
We take $\alpha_s(m_\phi)$ from~\cite{Bethke:2006ac}. Following~\cite{Grinstein:1988yu} we assume that the perturbative spectator model is valid at ${m_\phi > 2\gev}$. The dispersive analysis holds for $m_\phi\lesssim 1.3\,\gev$, where $\pi\pi$ and $KK$ dominate the hadronic decay rate. In the regime $m_\phi=1.3-2\gev$, significant corrections are expected. We will use the dispersive results up to $2\gev$, but include an additional contribution
\begin{equation}
 \Gamma_{4\pi,\eta\eta,\rho\rho,\dots} = C\,s_\theta^2\,m_\phi^3\beta_{2\pi}\,,
\end{equation}
to account for the increasing number of hadronic channels opening above the $4\pi$ threshold. The mass scaling is leaned upon the gluon channel. Setting $C=5.1\cdot 10^{-9}\gev^{-2}$, the hadronic decay rate transits smoothly into the rate of the spectator model at $m_\phi=2\gev$. In reality, peaks may occur due the further scalar resonances $f(1370)$, $f(1500)$, $f(1710)$. The strong increase of the hadronic decay rate around $\gev$, however, arises since $f_0(980)$ is narrow and located just below the kaon threshold to which it strongly couples~\cite{Truong:1989my}. A comparable situation does not seem to occur for the heavier scalar resonances and similar enhancements are, hence, not expected at higher mass. We may anticipate that the hadronic decay rates we obtain at $m_\phi=1.3-2\gev$ provide at least a valid order-of-magnitude estimate. In figure~\ref{fig:decayrates}, we depict the leptonic and hadronic decay rates of the light scalar below the $\bar{b}b$-threshold.

\section{Experimental Constraints and Future Sensitivities}\label{sec:constraints}

Experimental limits on light scalars as well as future sensitivities have been summarized various times, recently in~\cite{Alekhin:2015byh,Krnjaic:2015mbs,Flacke:2016szy,Evans:2017kti,Feng:2017vli,Evans:2017lvd}. These crucially depend on the decay properties of the scalar. In many instances, constraints with different assumptions on the hadronic decay rate have been combined. We will, therefore, reevaluate the existing limits on light scalars consistently using our new set of decay rates. Sensitivities of some important future searches will also be discussed. Our focus is on the mass window $m_\phi\simeq 0.01-10\gev$ accessible to accelerator probes.

\subsection{Rare Decays}
Light scalars can mediate rare meson decays. The most relevant processes include radiative $\Upsilon$-decays as well as flavor changing $B$ and $K$ meson decays (see figure~\ref{fig:raremeson}). The calculation of the corresponding branching ratios is summarized in appendix~\ref{sec:rare}.

\begin{figure}[h]
\begin{center}   
 \includegraphics[height=2.9cm,width=4.1cm]{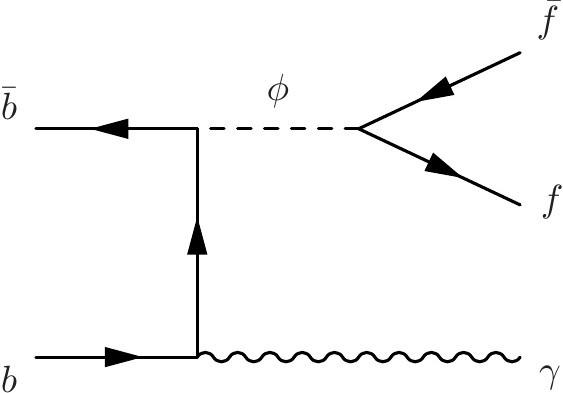}
 \includegraphics[height=2.9cm,width=4.1cm,trim={0 0 0 1.cm},clip]{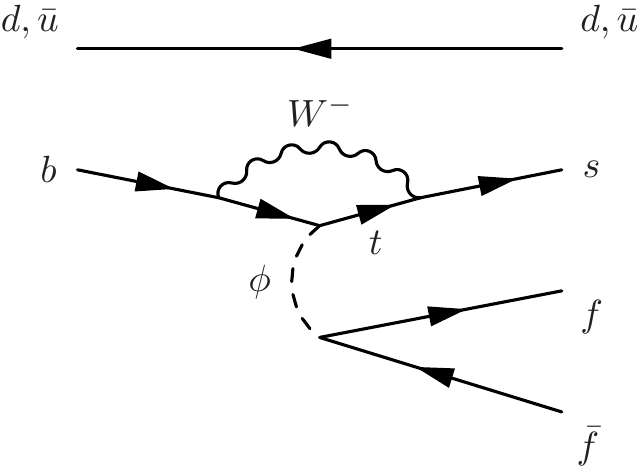}
\end{center}
\caption{Radiative $\Upsilon$ decays and flavor changing $B$ decays mediated by a light scalar.}
\label{fig:raremeson}
\end{figure}

BaBar has performed various searches for radiative $\Upsilon$ decays mediated by a light scalar. The most important channel is $\Upsilon\rightarrow \gamma + \text{jets}$ triggered by a hadronically decaying $\phi$~\cite{Lees:2011wb}. 

\begin{figure}[htp]
\begin{center}
\includegraphics[width=7.5cm,height=8cm]{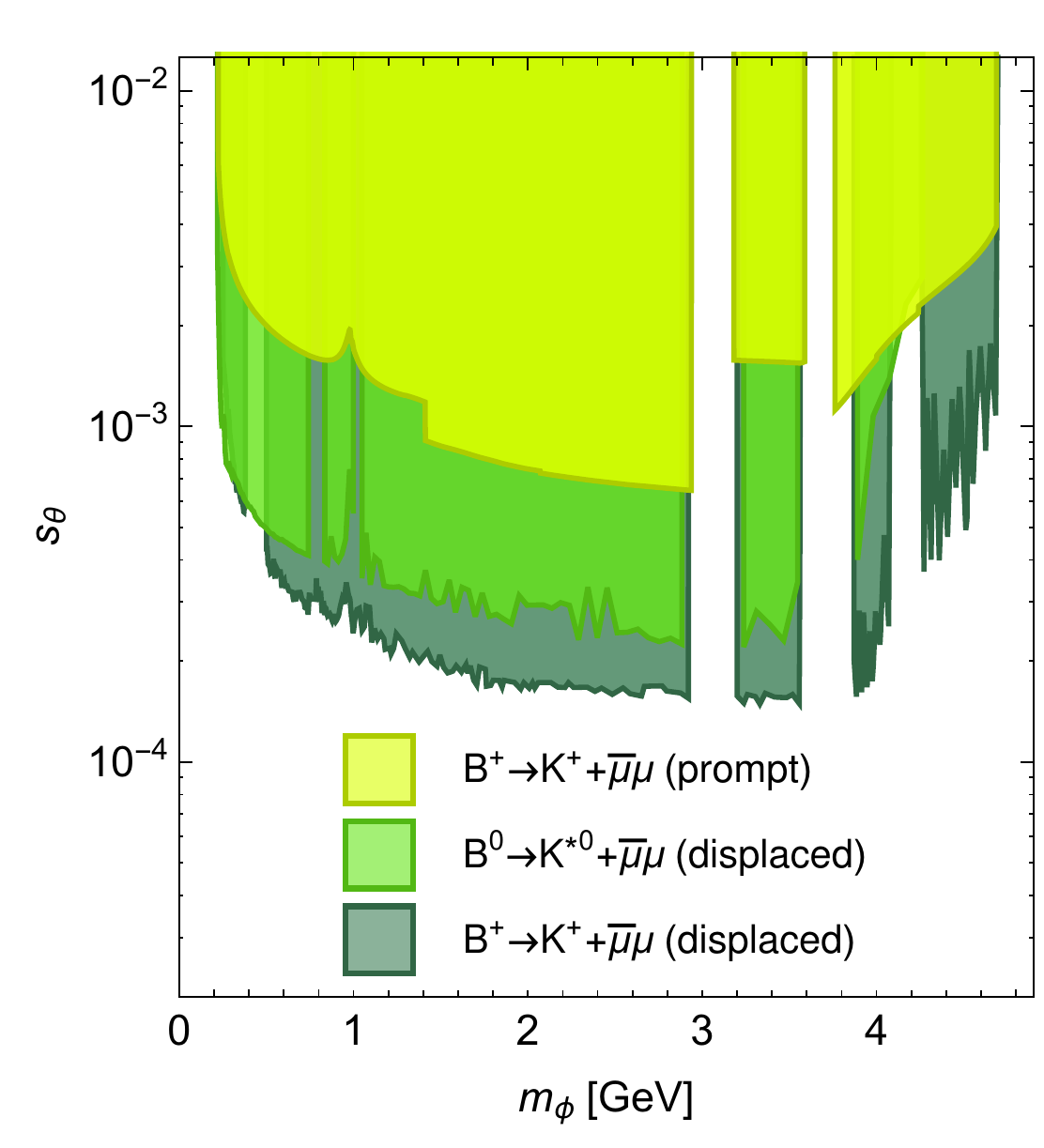} 
\end{center}
\caption{LHCb constraints on light scalars derived from rare $B$ decays.}
\label{fig:brare}
\end{figure}

Below the $B$ threshold, searches for semi-leptonic $B$ decays become relevant. LHCb measured the branching ratio $B^+\rightarrow K^+ +\bar{\mu}\mu$ in several bins of dilepton invariant mass~\cite{Aaij:2012vr}. The corresponding upper limit on the $\phi$-induced branching ratio in each bin is determined as in~\cite{Schmidt-Hoberg:2013hba}. It must be taken into account that LHCb triggered on prompt decays in this search. Following~\cite{Schmidt-Hoberg:2013hba}, we estimate that events with a (boosted) scalar decay length $d<d_{\text{max}}\simeq 5\:\text{mm}$ are reconstructed. This translates to an efficiency factor
\begin{equation}\label{eq:lhcbeff}
 \eta = \int\limits_0^\infty\! \D p_\phi \, f(p_\phi)\,\left( 1-e^{-m_\phi\Gamma_\phi d_{\text{max}}/p_\phi}\right)\,,
\end{equation}
where $f(p_\phi)$ denotes the momentum distribution of $\phi$ which is obtained with PYTHIA~\cite{Sjostrand:2014zea}.\footnote{We generated a large sample of $B$ mesons with PYTHIA and decayed each $B$ further to $\phi$ using the appropriate kinematics.} 	
LHCb has subsequently performed dedicated searches for light scalars with macroscopic decay lengths. In~\cite{Aaij:2015tna,Aaij:2016qsm} constraints on $\text{Br}_{B^0\rightarrow K^{*0}\phi}\times\text{Br}_{\phi\rightarrow\bar{\mu}\mu}$ and $\text{Br}_{B^+\rightarrow K^+\phi}\times\text{Br}_{\phi\rightarrow\bar{\mu}\mu}$ have been set as a function of the intermediate scalar mass and lifetime. We digitized the provided images and derived the corresponding constraints on $s_\theta$.\footnote{The case of a light scalar mixing with the Higgs has been covered explicitly in the two references. We, nevertheless, rederive the constraints on $s_\theta$ since a different set of scalar decay rates has been employed in~\cite{Aaij:2015tna,Aaij:2016qsm}.} As can be seen in figure~\ref{fig:brare}, the inclusion of displaced decays has significantly increased the LHCb sensitivity to light scalars in most of the mass range. A search for long-lived particles in $B$ decays was also performed by BaBar which looked for the inclusive process $B\rightarrow X_s  \phi$ with $\phi$ further decaying into leptons or hadrons~\cite{Lees:2015rxq}. The pion channel is most relevant since it excludes a small parameter region not covered by the previously mentioned LHCb searches.

Scalar masses of up to a few hundred MeV can be probed by rare kaon decays. We include the upper limit $\text{Br}_{K_L\rightarrow \pi^0+\bar{\mu}\mu}<3.8\cdot 10^{-10}$ stemming from the KTeV experiment~\cite{AlaviHarati:2000hs} in our analysis. Again, $\phi$-mediated processes only contribute to the rate if $\phi$ decays sufficiently promptly. Since KTeV is a fixed-target experiment, event reconstruction mostly depends on the transverse vertex location. Following~\cite{Dolan:2014ska} we assume that events with a (boosted) transverse scalar decay length below $4\:\text{mm}$ pass the trigger. The corresponding efficiency factor is calculated using~\eqref{eq:lhcbeff} with $p_\phi$ replaced by the transverse momentum. The distribution of transverse momenta is again determined with PYTHIA. 

Below the muon threshold, the scalar typically escapes detection due to its long lifetime. It still leaves a trace in the form of missing energy. The search for $K^+\rightarrow \pi^+ + \bar{\nu}\nu $ by E949 is used to set limits on $\text{Br}_{K^+\rightarrow \pi^+  \phi}$ as a function of the scalar mass and lifetime~\cite{Artamonov:2009sz}. In this case, visible decays of $\phi$ are vetoed, and the sensitivity increases with the lifetime of the scalar. We determine the corresponding exclusion in the $m_\phi$-$s_\theta$ plane.

Figure~\ref{fig:limits} shows that rare decays set the strongest constraints on light scalars over wide regions of the parameter space. Mixing angles down to $s_\theta=10^{-3}-10^{-4}$ are excluded for $m_\phi < m_B-m_K$ unless $m_\phi$ resides in the vicinity of the charmonium resonances $J/\psi$, $\psi(2S)$. The limits substantially degrade once scalar production in $B$ decays becomes kinematically inaccessible.

\subsection{Collider Searches}
At LEP, searches for Higgs bosons and Higgs-like scalars have been performed through the process $\bar{e}e\rightarrow Z^*\phi$. In the considered mass window, the strongest constraints are set by $L3$~\cite{Acciarri:1996um}. Strictly speaking, these apply to scalars which share the exact decay properties of a SM Higgs boson (at the considered mass). Since the mixing angle $s_\theta$ suppresses the couplings of $\phi$ compared to the Higgs one may worry that the longer decay length invalidates the bounds. This is not the case: for the range of $s_\theta\gtrsim 0.1$ covered by the search, a light SM Higgs and a light scalar would both decay mostly invisibly (on detector scales) below the muon threshold and visibly above. Even in the GeV range, the $L3$ analysis can be considered robust since it merely relies on the dominance of hadronic decay modes, while the particular enhancement of the pionic decay rate does not play a role. Above the $B$ meson mass, LEP still sets the strongest constraints on light scalars (see figure~\ref{fig:limits}). 

Turning to the LHC, light scalars are constrained by the search for spin-0 resonances in the dimuon channel. CMS and LHCb provided constraints on $\sigma_{pp\rightarrow\phi} \times \text{Br}_{\phi\rightarrow\bar{\mu}\mu}$ at $\sqrt{s}=7\tev$ and $8\tev$ respectively~\cite{Chatrchyan:2012am,Aaij:2018xpt} (see also~\cite{Haisch:2016hzu}). In the covered mass range $m_\phi=5.5-15\gev$, scalar production by $B$-meson decay is kinematically forbidden which makes gluon fusion the relevant process. We calculated the corresponding cross section with the tool SUSHI~1.6.1~\cite{Harlander:2012pb} in order to translate the limits into exclusions on $s_\theta$ (see figure~\ref{fig:limits}). Additional LHC constraints on light scalars arise from the non-observation of exotic Higgs decays. These shall not be considered in this work since they rely on the model-dependent Higgs-scalar coupling and, furthermore, only lead to subdominant exclusions in the considered mass range~\cite{Flacke:2016szy}. For proposed detector concepts (MATHUSLA, CODEX-b, FASER) which would increase the LHC sensitivity to light scalars, we refer to~\cite{Evans:2017lvd,Feng:2017vli,Evans:2017kti}.

\subsection{Beam Dump Experiments}
Beam dump experiments with detectors located $\mathcal{O}(100\:\text{m})$ away from the interaction point provide a sensitive laboratory to search for long-lived particles. Light scalars are most efficiently generated by $B$ and $K$ meson decays. For a proton beam impinging on a thick target which absorbs hadrons efficiently, the number of produced scalars can be estimated as
\begin{equation}
 N_\phi \simeq N_{\text{p.o.t.}}\! \left(n_B\,\text{Br}_{B\rightarrow X_s \phi} + \langle \gamma_K^{-1}\rangle \ell_H n_K\, \Gamma_{K\rightarrow \pi\phi}\right)
\end{equation}
with $N_{\text{p.o.t.}}$ denoting the number of protons on target. The multiplicities $n_B$, $n_K$ stand for the number of $B$, $K$ mesons created per incoming proton, $\langle\gamma_K^{-1}\rangle$ for the mean inverse kaon Lorentz factor. In the case of kaons only $K^\pm$ and $K_L$ should be considered since $K_S$ has a suppressed decay rate to scalars (see appendix~\ref{sec:rarek}). Different from $B$ mesons, most kaons are absorbed in the target since their decay length exceeds the hadronic absorption length $\ell_H$.\footnote{While the decay length differs substantially between $K^\pm$ and $K_L$, $\Gamma_K^{-1}\gg \ell_H$ holds for both species.} The above approximation neglects kaon regeneration by secondary interactions. Furthermore, it assumes that the number of kaons escaping the target is negligible (as is valid for a target with a thickness of several $\ell_H$). 

The probability $\mathcal{P}_\phi$ that a scalar with three-momentum $\mathbf{p}_\phi$ leaves a signal in the detector reads
\begin{equation}
 \mathcal{P}_\phi = \int\limits_{d_1}^{d_2} \!\D z \:\frac{\eta_{\text{geom}}\,\eta_{\text{rec}}\:m_\phi\, \Gamma_\phi}{p_\phi} 
\: e^{-m_\phi \Gamma_\phi z/p_\phi}\,.
\end{equation}
The decay vertex $z$ of the scalar needs to be located within the distance $d_1-d_2$ from the target to be detected. The geometric efficiency $\eta_{\text{geom}}$ accounts for the probability that the decay products of $\phi$ pass through the detector. It depends on the angular coverage of the detector and varies with the scalar's momentum and the location $z$ of the decay vertex. The factor $\eta_{\text{rec}}$ is the reconstruction efficiency for final states of a certain type. In order to determine the total number of events, we need to integrate the product $N_\phi \mathcal{P}_\phi$ over the momentum distribution of $\phi$. The latter is again determined with PYTHIA by creating large samples of $B$ and $K$ mesons which are then decayed further to scalars. Kaon events are properly weighted to account for the fact that highly boosted kaons are more likely to be absorbed due to their longer decay length. The geometric efficiency is determined from the momentum spectrum of $\phi$ by decaying the scalars and selecting events with all final states passing through the detector. 

We consider the CHARM beam dump (which operated in the 1980s), the upcoming run of NA62 in dump mode and the planned SHiP experiment. All three detectors have been/ will be located at the CERN SPS and employ a 400~GeV proton beam.\footnote{A search for long-lived scalars could potentially also be performed at the Fermilab SeaQuest Experiment after minor modifications of the setup~\cite{Berlin:2018pwi}.} The meson multiplicities are estimated as $n_B\simeq 3.2\cdot 10^{-7}$~\cite{Anelli:2015pba} and $n_K\simeq 0.9$~\cite{Antinucci:1972ib}.\footnote{We extracted $n_{K^\pm}=0.62$ from~\cite{Antinucci:1972ib} and estimated $n_{K_L}\simeq 0.28$ by taking the $K_L/K^\pm$ ratio from PYTHIA.}
The target materials copper (CHARM, NA62) and molybdenum (SHiP) share a hadronic absorption length $\ell_H\simeq 15.3\:\text{cm}$~\cite{Groom:2017}. Locations and coverage of the detectors are described in~\cite{Bergsma:1985qz,NA62:2017rwk,Anelli:2015pba}. CHARM is sensitive to leptonic final states with efficiency 0.5~\cite{Bergsma:1985qz}. SHiP and NA62 should be sensitive to all sorts of final states with $\eta_{\text{rec}}=0.4$ ($\eta_{\text{rec}}=0.7$)
below (above) the two-muon threshold for SHiP~\cite{Anelli:2015pba} and $\eta_{\text{rec}}\simeq1$ for NA62~\cite{Lanfranchi:2017wzl}. We summarize the luminosities, locations of the decay volumes and mean geometric efficiencies $\overline{\eta}_{\text{geom}}$ (for detection of $B$- and $K$-induced scalars) in table~\ref{tab:beamdumps}.\footnote{The mean geometric efficiency $\overline{\eta}_{\text{geom}}$ was derived by averaging $\eta_{\text{geom}}$ over the momentum distribution and the location of the decay vertex within $d_1-d_2$. The stated ranges are obtained by varying the scalar mass between $0.01\gev$ and $m_B-m_K$.} SHiP will be a factor $\mathcal{O}(10^4)$ more sensitive compared to its predecessors due to the larger beam intensity and the better detector coverage.
\begin{table}[t]
\begin{center}
\begin{tabular}{|cccc|}
\hline
 &   &  &     \\[-4mm]
  & $\text{N}_{\text{p.o.t}}$ & $d_1-d_2$~[m] & $\overline{\eta}_{\text{geom}}$  \\[1mm]
\hline \hline
  &   & &\\[-3mm]
 CHARM$\;\;$    &  $2.4\cdot 10^{18}$ & $480-515$ & \parbox{3cm}{$0.001-0.002\;(K)$\\$0.002-0.01\phantom{0}\;(B)$} \\[3mm]
 NA62    &  $10^{18}$ & $95-160$ & \parbox{3cm}{$0.002-0.005\;(K)$\\$0.002-0.02\phantom{0}\;(B)$} \\[3mm]
 SHiP    &  $2\cdot 10^{20}$ & $69-120$ & \parbox{3cm}{$\phantom{0}0.05-0.08\phantom{0}\;(K)$\\$\phantom{0}0.2\phantom{0}-0.5\phantom{0}\phantom{0}\;(B)$}  \\[2mm]
 \hline
\end{tabular}
\end{center}
\caption{Comparison between the CHARM, NA62 and SHiP beam dump experiments.}
\label{tab:beamdumps}
\end{table}

\begin{figure*}[htp]
\begin{center}
 \includegraphics[width=17cm]{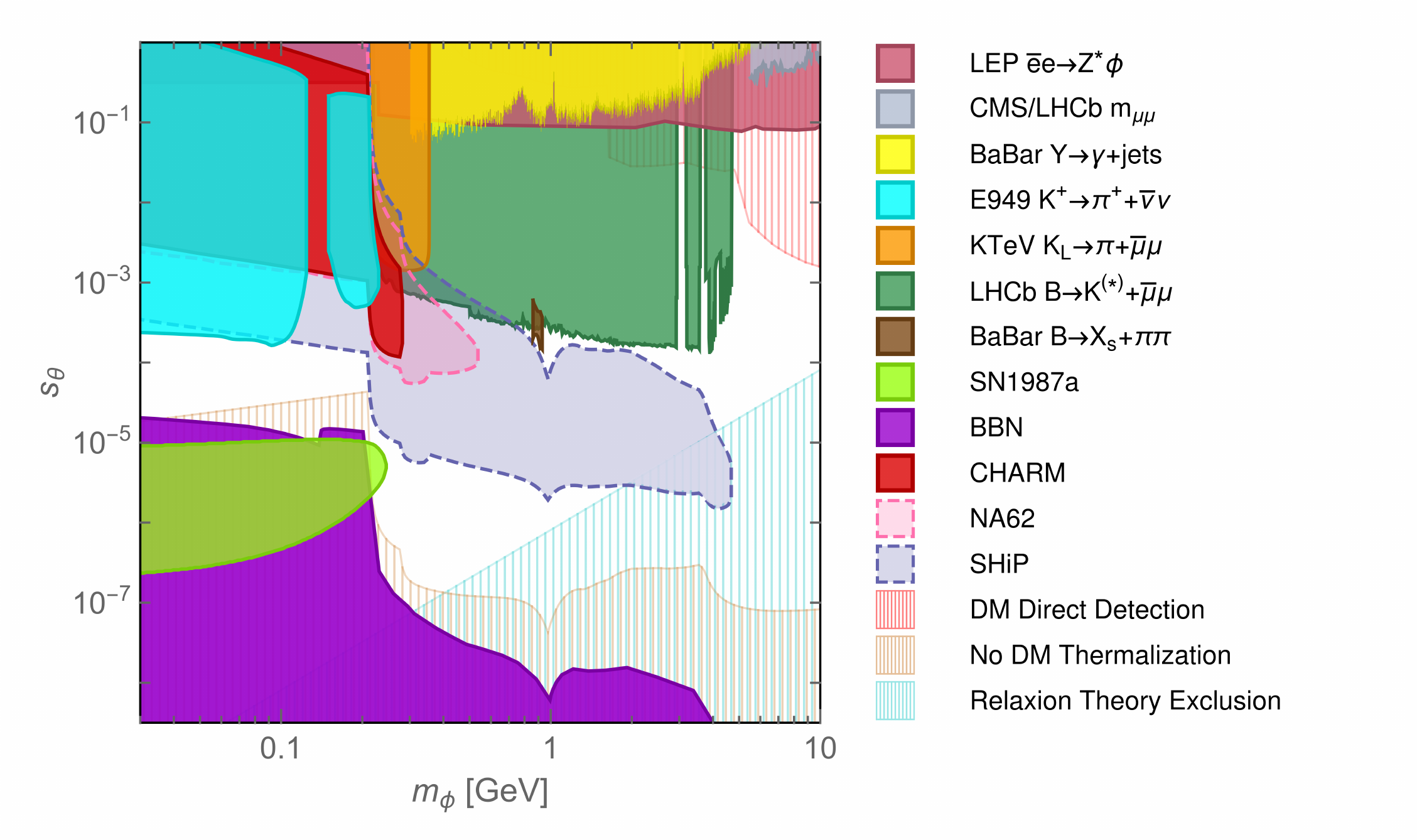}
\end{center}
\caption{Constraints on light scalars mixing with the Higgs. The filled regions with solid boundaries correspond to model-independent constraints. Sensitivity projections are indicated by the dashed boundary. The hatched regions refer to model-dependent exclusions which apply to the relaxion model (cyan) and the dark matter model (red, ocher) discussed in section~\ref{sec:models}.\\[3mm]} 
\label{fig:limits}
\end{figure*}

CHARM did not observe any signal events which translates to an upper limit of 3 expected events (at $95\%$~confidence level). The corresponding exclusion on light scalars reaches down to $s_\theta \sim 10^{-4}$ (see figure~\ref{fig:limits}). We note that the CHARM constraint obtained by us is substantially weaker than in previous evaluations~\cite{Clarke:2013aya,Dolan:2014ska,Alekhin:2015byh,Krnjaic:2015mbs,Flacke:2016szy,Evans:2017kti,Feng:2017vli,Evans:2017lvd}. We believe that in these references, kaon absorption in the thick copper target -- which drastically reduces $N_\phi$ from kaon decays -- has been neglected. 

\begin{figure}[t]
\begin{center}
\includegraphics[width=7.8cm]{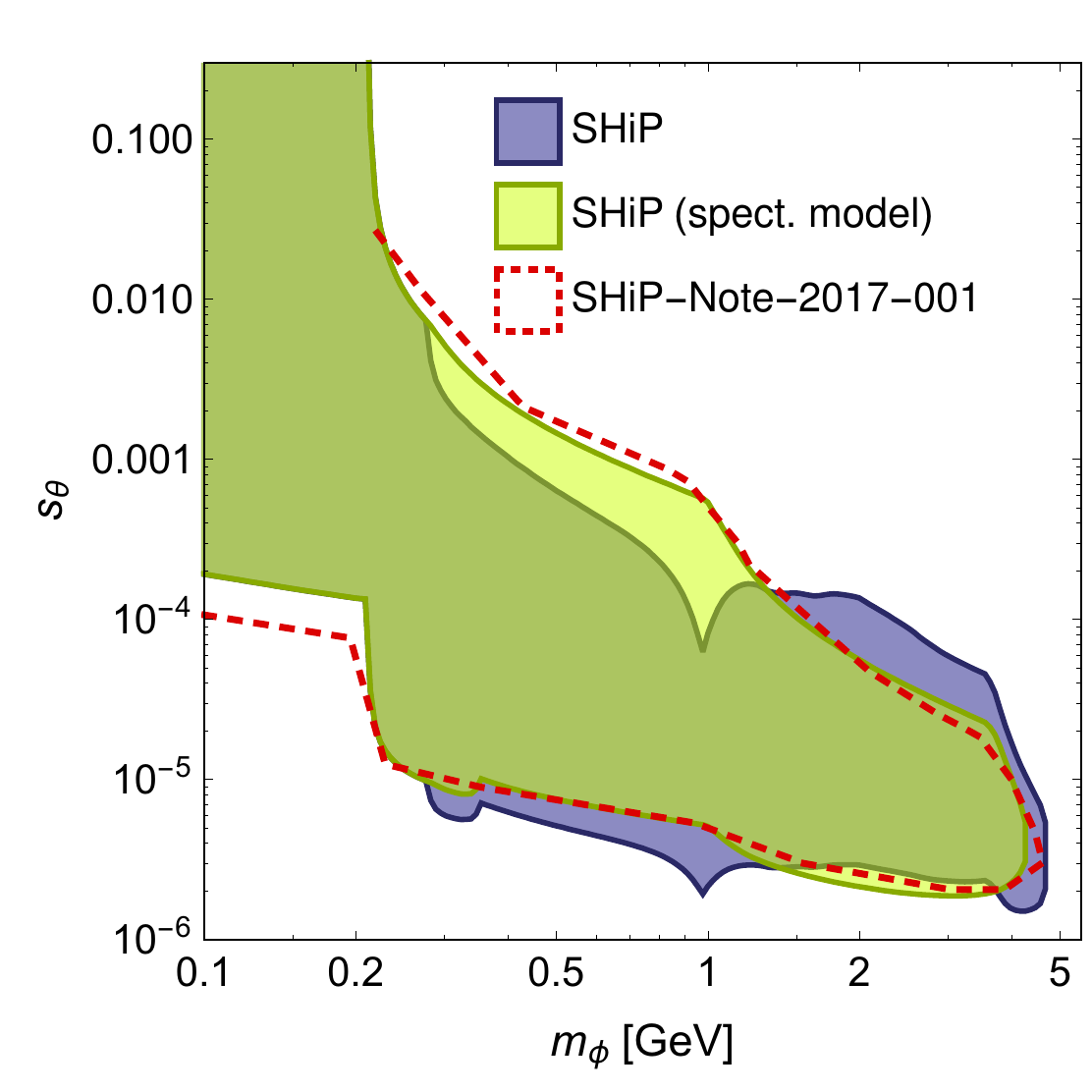} 
\end{center}
\caption{SHiP sensitivity to light scalars found in this work compared to~\cite{Lanfranchi:2017}. The blue shaded region is obtained for the scalar decay rates derived in section~\ref{sec:scalardecay} and represents our preferred estimate. The yellow region is obtained if we treat the scalar decay in the perturbative spectator model.}
\label{fig:shipcomparison}
\end{figure}

Sensitivity projections for NA62 and SHiP in figure~\ref{fig:limits} again correspond to 3 events. They should be considered as optimistic since a negligible background level was assumed. While the number of produced scalars in NA62 is similar as in CHARM, NA62 is sensitive to higher masses since it can reconstruct pion final states. SHiP will cover a huge parameter region not previously accessible to any experiment. For SHiP and NA62, we again find deviations from the semi-official sensitivity estimates~\cite{Lanfranchi:2017,Lanfranchi:2017wzl} (see figure~\ref{fig:shipcomparison}). 

In this case, the discrepancy can be traced back to the assumptions on the scalar decay rates. While we relied on a dispersive analysis in the non-perturbative QCD regime (see section~\ref{sec:scalardecay}), the perturbative spectator model has been employed in~\cite{Lanfranchi:2017,Lanfranchi:2017wzl}. In figure~\ref{fig:shipcomparison} it can be seen that our sensitivity estimate approximately reproduces the SHiP projection from~\cite{Lanfranchi:2017} if we also switch to the spectator model. The same observation is made in the case of NA62. We emphasize, however, that our dispersive analysis provides a much more realistic description of the scalar decay properties in the GeV range compared to the spectator model.

We finally comment that the sensitivity of NA62 to light scalars could be significantly improved: the present estimate refers to the experiment running in dump mode. This means that the beryllium target is lifted and the collimator is closed such that it acts as dump for the proton beam. The disadvantage of this layout is that most produced kaons are absorbed in the thick collimator before they can decay. It appears preferential to leave the (thin) beryllium target in the beam line and keep the collimator closed. The latter would then still filter hadronic backgrounds. But since it is located $20\:\text{m}$ downstream the target, a significant fraction of the kaons created in the target could decay before reaching the collimator. This would increase the number of light scalars from kaon decay by a factor 10-100 compared to dump mode.

\subsection{Cosmology and Astrophysics}
Light scalars can also be constrained by requiring that they do not spoil the cosmological evolution. In the hot early universe, the light scalars are copiously produced in the thermal bath. Due to their small coupling to SM matter, their freeze-out abundance is significant. If their decay happens after the onset of primordial nucleosynthesis (BBN), the hadronic energy injection would have spoiled the light element abundances. The resulting upper limit on the scalar lifetime ranges from $1/100\s-1\s$ in the considered mass range~\cite{Fradette:2017sdd}.\footnote{The constraint mildly depends on the (model-dependent) Higgs-scalar coupling and was shown for three different choices in~\cite{Fradette:2017sdd}. To be conservative we used the weakest of the three constraints at each mass.} 
It was converted to a constraint on $s_\theta$ by using the decay rate from figure~\ref{fig:decayrates}.

Finally, astrophysical processes can be affected by light scalars. Most importantly, scalar emission could carry away significant amounts of energy in supernova explosions~\cite{Ellis:1987pk,Turner:1987by}. This would lead to a shortening of the neutrino pulse which is constrained by observations of SN1987a. We determine the corresponding exclusions on light scalars following the treatment described in~\cite{Krnjaic:2015mbs,Evans:2017kti}. 

While accelerator searches exclude large mixing angles, cosmology constrains $s_\theta$ from below (see figure~\ref{fig:limits}). For $m_\phi\lesssim 5\gev$, a window of $s_\theta\sim10^{-3}-10^{-5}$ and $s_\theta\sim10^{-4}-10^{-8}$ remains viable below and above the two-muon threshold respectively. In models, where the light scalar is identified with the relaxion (section~\ref{sec:relaxion}) or the mediator connecting to dark matter (section~\ref{sec:darkmatter}), additional constraints apply which close parts of this window. Nevertheless, there remains an exciting discovery potential for the next generation of experiments.

\section{Conclusion}
We have reinvestigated the decay properties of a light scalar boson mixing with the Higgs. A special focus was placed on the mass range $m_\phi\simeq 0.5-2\gev$ in which hadronic decay modes are affected by strong final state interactions. We performed a new dispersive analysis and derived the decay rates of the scalar to pions and kaons. These were confronted with two earlier evaluations by Donoghue et al.~\cite{Donoghue:1990xh} and Truong \& Willey~\cite{Truong:1989my} which are inconsistent with one another. Our result confirms the calculation of Donoghue et al.\ to within $\mathcal{O}(1)$ precision. The remaining difference can be explained by our updated input of pion-kaon phase shift data. We also showed that Truong \& Willey's result is brought into qualitative agreement with our calculation, once a sign error in their $T$-matrix parameterization is corrected. By matching the dispersive calculation to the perturbative spectator model at higher mass, we obtained a realistic estimate of scalar decay rates over the full mass range (figure~\ref{fig:decayrates}). We also provided the hadronic form factors which allow to generalize our result to non-universally coupled light scalars (figure~\ref{fig:formfactors}).

Finally, we rederived the accelerator-, cosmological and theoretical constraints on light scalars in the MeV-GeV mass window (figure~\ref{fig:limits}). We covered the model-independent case as well some of the most prominent explicit models with light scalars. Sensitivity projections for future key searches were also provided. The strongest deviations compared to previous evaluations occur for beam dump experiments. In the case of CHARM, previous exclusions were too restrictive since they had neglected kaon absorption in the target. In addition, our new-found decay rates strongly impact the sensitivity window of beam dumps by affecting the decay length of light scalars.

\appendix

\section{Scalar in Rare Decays}\label{sec:rare}

\subsection{Radiative $\Upsilon$ decays}
A light scalar can emerge in the radiative decay $\Upsilon \rightarrow \gamma \, \phi$ and induce a meson or lepton pair~\cite{Wilczek:1977zn}. It is convenient to express the corresponding branching ratio in the form
\begin{equation}\label{eq:upsilon}
 \frac{\text{Br}_{\Upsilon\rightarrow \gamma\,\phi}}{\text{Br}_{\Upsilon\rightarrow \bar{e} e }}=\frac{s_\theta^2\, G_F m_b^2}{\sqrt{2}\pi\alpha}\,\mathcal{F}\,\Big(
1-\frac{m_\phi^2}{m^2_\Upsilon}\Big)\;,
\end{equation}
where $\alpha$ is the Sommerfeld constant and $\mathcal{F}$ a correction function taken from~\cite{Ellis:1985yb}. It accounts for higher order QCD processes~\cite{Vysotsky:1980cz,Nason:1986tr} as well as bound state effects appearing close to the kinematic endpoint~\cite{Haber:1978jt,Ellis:1979jy}.

\subsection{Rare $B$ Decays}
The scalar appears in an effective flavor violating coupling $\phi$-$s$-$b$. By integrating out the $W$-$t$-loop one obtains~\cite{Batell:2009jf}
\begin{align}
 \mathcal{L}_{\phi sb}&=g_{\phi sb} \phi\, \bar{s}_L b_R +\text{h.c.}\,,\nonumber\\
 g_{\phi sb}&=\frac{s_\theta\,m_b}{v}\,\frac{3\sqrt{2}\,  G_F \,m_t^2\, V_{ts}^* V_{tb}}{16\pi^2}\,,\label{eq:effcoupling}
\end{align}
where $V_{ts}$ and $ V_{tb}$ denote the CKM matrix elements. The above Lagrangian triggers the decay $B \rightarrow K^{(*)}\phi$ for which the rate reads
\begin{equation}
\label{eq:BKphi}
\Gamma_{B \rightarrow K^{(*)}\phi} =\left|g_{\phi sb}\right|^2\,\left|\langle K^{(*)}|\bar{s}_L b_R|B\rangle\right|^2 \frac{\lambda_{B, K^{(*)}\phi}^{1/2}}{16\pi\,m_B}\,,
\end{equation}
where we introduced
\begin{equation}
\lambda_{x,yz}=\frac{m_x^2-(m_y-m_z)^2}{m_x^2}\frac{m_x^2-(m_y+m_z)^2}{m_x^2}\,.
\end{equation}
The matrix elements can be approximated as~\cite{Ball:2004ye,Ball:2004rg}
\begin{align}
 |\langle K^*|\bar{s}_L b_R|B\rangle|^2&=\frac{1}{4}\frac{m_B^4\;\lambda_{B, K^{(*)}\phi}}{(m_b+m_s)^2}\,{A_{K^*}^2}\,,\nonumber\\
 |\langle K|\bar{s}_L b_R|B\rangle|^2 &=\frac{1}{4}\frac{(m_B^2-m_K^2)^2}{(m_b-m_s)^2}\,{f_{K}^2}\,
\end{align}
with
\begin{align} 
A_{K^*}&= \frac{1.36}{1-q^2/27.9\gev^2}-\frac{0.99}{1-q^2/36.8\gev^2}\,,\nonumber\\
f_{K}&= \frac{0.33}{1-q^2/37.5\gev^2}\,.
\end{align}
The transferred momentum is set to $q^2=m_\phi^2$. In the case of $K^*$ we already took the sum over polarizations.

For cases where the nature of the strange particle(s) in the final state is not of relevance, one can define the inclusive decay rate $B \rightarrow X_s\,\phi$. The spectator model predicts~\cite{Willey:1982mc}
\begin{equation}\label{eq:inclusive}
\Gamma_{B \rightarrow X_s\phi} =\left|g_{\phi sb}\right|^2\frac{(m_B^2-m_\phi^2)^2}{32\pi\,m_B^3}\,.
\end{equation}
This estimate is not valid close to the kinematic endpoint, where the spectator model breaks down. In this regime, the inclusive rate should, however, converge towards $\Gamma_{B \rightarrow K\phi}$ since this is the only available final state. In order to obtain a smooth function with the correct asymptotic behavior, we use~\eqref{eq:inclusive} for $m_\phi<4.7\gev$ and set $\Gamma_{B \rightarrow X_s\phi} =\Gamma_{B \rightarrow K\phi}$ above.

\subsection{Rare $K$ Decays}\label{sec:rarek}
The scalar can also induce rare decays of lighter mesons, for instance $K\rightarrow \pi \,\phi$. The corresponding decay rate is again dominated by the $W$-$t$-loop. One finds\footnote{We neglect subleading contributions due to scalar brems-strahlung and the charm loop which would amount to a correction $\lesssim 10\%$~\cite{Leutwyler:1989xj}.}~\cite{Willey:1982ti}
\begin{equation}
\label{eq:Kpiphi}
\Gamma_{K^{\pm} \rightarrow \pi^{\pm}\phi} \simeq
\left|g_{\phi ds}\right|^2\,\left|\langle \pi|\bar{d}_L s_R|K\rangle\right|^2 \frac{\lambda_{K, \pi\phi}^{1/2}}{16\pi\,m_K}\,,
\end{equation}
and $\Gamma_{K_L \rightarrow \pi^0\phi}\simeq\Gamma_{K^{\pm} \rightarrow \pi^{\pm}\phi}$. The effective coupling $g_{\phi ds}$ is obtained from~\eqref{eq:effcoupling} by the replacement $(b,s)\rightarrow (s,d)$. The matrix element reads~\cite{Kamenik:2011vy}
\begin{equation}
 \left|\langle \pi|\bar{d}_L s_R|K\rangle\right| \simeq \frac{1}{2}\frac{(m_K^2-m_\pi^2)}{m_s-m_d}\,.
\end{equation}
Since the corresponding rate for the $K_S$ decays is proportional to the small CP violating phase in the CKM matrix, it suffers a stronger suppression~\cite{Leutwyler:1989xj}.

\section*{Acknowledgements}
I would like to thank Kai Schmidt-Hoberg, Felix Kahlh\"ofer, Katherine Freese, Luca Visinelli and Sebastian Baum for helpful discussions.

\bibliography{btex}
\bibliographystyle{ArXiv}

\end{document}